# Mid-Cycle Observations of CR Boo and Estimation of the System's Parameters

Daniela Boneva [1,*], Svetlana Boeva [2], Yanko Nikolov [2], Zorica Cvetković [3] and Radoslav Zamanov [2]

[1]   Space Research and Technology Institute, Bulgarian Academy of Sciences, 1113 Sofia, Bulgaria
[2]   Institute of Astronomy and National Astronomical Observatory, Bulgarian Academy of Sciences, 1784 Sofia, Bulgaria; sboeva@astro.bas.bg (S.B.); ynikolov@astro.bas.bg (Y.N.); rkz@astro.bas.bg (R.Z.)
[3]   Astronomical Observatory, 11060 Belgrade, Serbia; zorica@aob.rs
*   Correspondence: danvasan@space.bas.bg

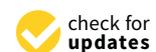

**Abstract:** We present observations (with NAO Rozhen and AS Vidojevica telescopes) of the AM Canum Venaticorum (AM CVn) binary star CR Bootis (CR Boo) in the UBV bands. The data were obtained in two nights in July 2019, when the V band brightness was in the range of 16.1–17.0 mag. In both nights, a variability for a period of 25 ± 1 min and amplitude of about 0.2 magnitudes was visible. These brightness variations are most likely indications of "humps". During our observational time, they appear for a period similar to the CR Boo orbital period. A possible reason of their origin is the phase rotation of the bright spot, placed in the contact point of the infalling matter and the outer disc edge. We estimated some of the parameters of the binary system, on the base of the observational data.

**Keywords:** astronomy; binary stars; AM CVn stars: CR Boo

## 1. Introduction and Object Details

CR Boo is a member of the AM CVn stars group. The AM CVn stars are short-period binary stars in which a white dwarf accretes helium-rich material from a low-mass donor star. Their orbital periods vary between 5 and 65 min [1,2].

CR Boo was discovered in 1986 by Palomar Green [3] and catalogued as PG 1346+082. The first observations of Wood et al. (1987) [4] showed a brightness variability with a V band magnitude in the range of 13.0–18.0 mag. The system displays spectroscopic variations—broad, shallow HeI absorption lines at maximum, and a weak emission in HeI 4471 at minimum light [4]. The average orbital period of CR Boo is estimated as ~1471.3 s to ~24.5 min or 0.0170290 days [5,6]. The masses of two components were estimated to be as follows: $M_1 = 0.7–1.1\ M_\odot$ for the mass of the primary star and $M_2 = 0.044–0.09$ for the mass of the secondary star [2,7].

The short orbital period, photometry and spectroscopy define CR Boo as an interacting double white dwarf object, in which the white dwarf primary accretes from the helium white dwarf companion [8–11].

Solheim has made a classification of AM CVn objects [2], dividing them in four groups, which depends on their orbital periods and disc's properties. According to its orbital period, CR Boo can be associated to the 3rd group, where the objects have an orbital period in a range of 20 < P < 40 min, with a variable size of the discs, producing outbursts or occasional super outbursts.

The AM CVn stars exhibit brightness variability, usually in the range of 2–4 magnitudes at optical bands [6,11–13]. Among the AM CVn stars, CR Boo has one of the best observational behaviors, because of the large-scale amplitude variations in brightness, from days to months, and it is easy to observe. CR Boo is categorized in the group of outburst systems [11,14], with a large amplitude variability of > 1 mag.

Two quiescent states of CR Boo have been observed: faint, with regular super outbursts [6,15], and bright, with frequent outburst activity [15,16]. During the faint state, the produced high outburst frequency is in a super cycle of about ~46 days [12,17]. It is similar to SU Uma type dwarf novae—a class of cataclysmic variables (CVs) [18]—where normal outbursts and super outbursts are observed; usually, both are with more than 1 amplitude variation in magnitude, while the latter are long-lasting bursts, from days to weeks.

Short-period, low-magnitude brightness variations could be observed during the outburst or super outburst states, or in quiescent states. Recognized as "humps" and "superhumps" [6,11], the two effects are clearly distinguished by Osaki and Meyer [19] and explained by Warner [18]. The "humps", or "orbital humps", are observed in the quiescence state of the cataclysmic variables and AM CVn stars. They appear for a period similar to the binary orbital period. The "superhumps" can be observed during the active (or outburst) state and their period is a few percent longer than the binary period.

In the next two sections, we report our observational results of CR Boo, calculate some of the system parameters and shortly discuss the observational effects at the end of Section 3.

## 2. Observational Details and Results

In this work, we report data of CR Boo from two nights: 2019/07/01 and 2019/07/05. The observations were obtained with the 2.0 m telescope of the National Astronomical Observatory (NAO) Rozhen, Bulgaria, the 50/70 cm Schmidt telescope (hereafter Sh) of NAO Rozhen and 1.4m AS (Astronomical Station) Vidojevica (hereafter Vid) telescope, Serbia. On the night of July 1st, the 2.0 m telescope was used, in the V band, equipped with a CCD camera VersArray, and the exposure was 60 s. On the night of July 5th, three different telescopes were used: the 2.0 m telescope for the U and V bands, with FoReRo—a 2-channel focal reductor. The exposure for U was 300 s and for V 60 s. The B filter was applied at the 50/70 cm Schmidt telescope with CCD camera FLI and the exposure was 120 s. Data in the B filter were also obtained with a 1.4 m telescope at AS Vidojevica, equipped with the CCD camera Andor. There were two continued exposures of 30 s and 60 s.

The data reduction was performed with standard IRAF tools [20] for processing of CCD images and aperture photometry. We used circular apertures with a size 4 arcsec, inner radius of sky annulus 5 arcsec and outer radius of the sky annulus 10 arcsec. Three comparison stars were used: TYC 900-1176-1, with V = 11.43 and B = 13.041; TYC 900-1047-1, with V = 12.076 and B = 12.552; and USNOA2 0975-07298478, with V = 13.739 and B = 14.991. The basic photometric standards from AAVSO—st 114 and st 121, as well as APASS 9—were applied. The estimated errors for each band were as follows: 0.01 in U and 0.02 in V; in UBV, 0.01–0.02 (2 m Rozhen telescope); 0.05–0.10 in B (1.4 m Vidojevica); and 0.05–0.10 in B (50/70cm Schmidt). IRAF tools were used to determine the uncertainties in the photometric data.

During the time of our observations the star was in-between outbursts, or probably before the high state, according to the super-cycle period of CR Boo of ~46 days [12,17] and comparing to AAVSO light curves [21]. We called this the "mid-cycle" observations.

On the first night of observations, 1 July, the star magnitude was 16.1−16.3 in V, with 0.2 amplitude variations and a periodicity of 0.015–0.017 days or 21.6–24.4 min (Figure 1). On the second night of observations, 5 July, the brightness decreased with ~0.7 magnitudes in its average magnitude values in V (Figure 2) compared to the 1st night, and its magnitude was 16.78–16.98 in V. The variability was again 0.1−0.2, with a periodicity of 0.015–0.018 days or 21.6–25.9 min. To analyze the periodicity of these brightness variations, we used the PDM (Phase Dispersion Minimization) method by Stellingwerf [22]. The average period of the measured amplitude variations is approximately the same as the orbital period Porb of 24.5 min, obtained in [5,6].

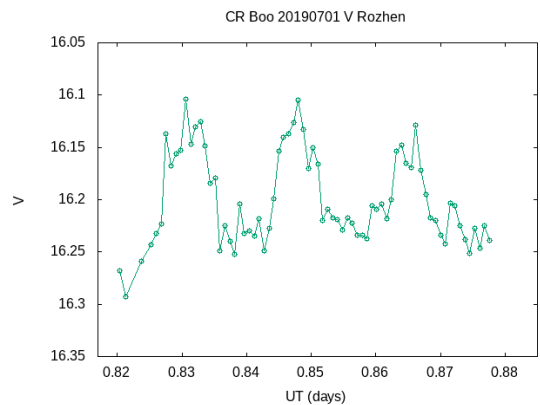

**Figure 1.** Light curves of CR Boo in the V band on 1 July 2019. The star's magnitude is 16.1–16.3 in V, with 0.2 amplitude variations and a periodicity of ~21.6–24.4 min. The data were obtained with a 2 m Rozhen telescope.

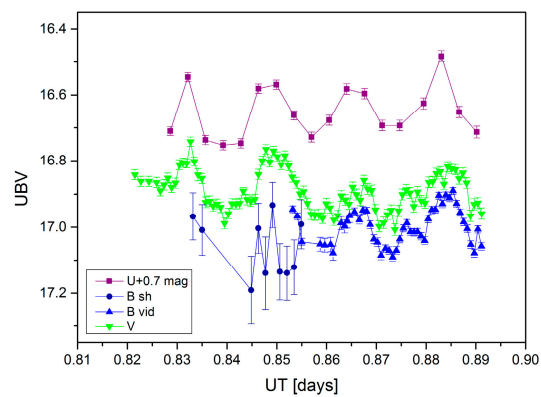

**Figure 2.** Light curves of CR Boo in the UBV bands on 5 July 2019. The star's magnitude is 16.78–16.98 in V, with 0.1–0.2 amplitude variations and periodicity of ~21.6–25.9 min. The brightness decreases with 0.7 magnitudes in comparison with the first night. To fit the U band light curve into the frame, these 0.7 magnitudes were added to the observational data (U + 0.7). The individual error bars are marked on each light curve.

Occasionally, a temporary fading that lasted a day was observed both during the quiescent period or even the outburst period of CR Boo, as reported in other works [6,11]. The fading in brightness with a 0.7 mag in V, observed on 5 July, is a manifestation of a similar peculiar variability, which happens through the mid-cycle in our case. The data obtained in this second night of our observations could be one more clue in support of such a specific behavior of this object.

Observations in the U-band, obtained in the second night (Figure 2), imply that the star probably has an active accretion disc. The magnitude ranges from 15.8 to 16.01, again with an average amplitude variation of ~0.2.

From the U-band data and by using the calculated luminosity Lu, we could further measure the mass accretion rate and estimate the temperature of the accretion disc. The relation between the U band emission, particularly the luminosity, and the accretion processes is detailed in [23,24].

Our observational results of CR Boo are in agreement with the star's magnitude range of 13.5–17.5 in V, as obtained in [4,5,14,25].

## 3. Calculations of the System Parameters and Observational Effects

Defining the properties of the star system are connected to the system parameters. For this study, we are taking into account the masses of two components and the orbital period, calculating the radii of the primary and secondary stars and the orbital separation. The ranges of masses $M_1$ and $M_2$ are already given in Section 1. We carefully select the values within these ranges, since the values of the masses reflect on the measurement of the orbital separation $a$ and on the radii of the components (see Equations (1–3)). We assume $M_1 = 0.80\ M_\odot$ and $M_2 = 0.07\ M_\odot$, which is close to the value 0.062 in [8]. Then, for the mass ratio we have q = 0.087, which is almost equal to the initially received value ~0.085 in [7] and the same as one of the mass ratio values in [6], and the total mass is M = $M_1 + M_2 = 0.87 M_\odot$. We use these values in further calculations. The orbital separation $a$ can be obtained by applying the Kepler's 3rd law:

$$a = \left( \frac{G(M_1 + M_2)M_\odot P^2}{4\pi^2} \right)^{\frac{1}{3}} \tag{1}$$

Then, from the calculations, we have $a = 0.266\ R_\odot = 1.84 \times 10^{10}$ cm. The orbital separation depends not only on the total mass and orbital period, it varies upon the mass transfer rate and mass ratio as well. When mass transfer is in progress between the two components, the mass ratio is getting to decrease and their orbital separation increases [26]. The high mass transfer is usually proper for the objects with short orbital periods, as is the period estimated for CR Boo.

The radius of the primary star $R_1$ is calculated using Eggleton's mass–radius relation [26,27]:

$$\frac{R_1}{R_\odot} = 0.0114 \left[ \left( \frac{M_1}{M_{Ch}} \right)^{-2/3} - \left( \frac{M_1}{M_{Ch}} \right)^{2/3} \right]^{1/2} \times \left[ 1 + 3.5 \left( \frac{M_1}{M_p} \right)^{-2/3} + \left( \frac{M_1}{M_p} \right)^{-1} \right]^{-2/3} \tag{2}$$

where $M_{Ch} = 1.4\ M_\odot$ is the Chandrasekhar mass; and $M_p = 0.00057\ M_\odot$ is a constant. We derive for $R_1 = 0.012\ R_\odot = 8.34 \times 10^8$ cm. This equation is applicable for the mass range $0 < M_{(1,2)} < M_{Ch}$.

Since, in our case, it is accepted that the mass transfer has already started, then the secondary star should have filled-up its Roche lobe, and its radius is approximately equal to the Roche lobe $R_L$, $R_2 \approx R_L$. To calculate $R_2$, the equation of Eggleton [28] is suitable:

$$\frac{R_2}{a} = \frac{0.49 q^{\frac{2}{3}}}{0.6 q^{\frac{2}{3}} + ln\left(1 + q^{\frac{1}{3}}\right)} \tag{3}$$

It gives $R_2 = 0.0526\ R_\odot = 3.6 \times 10^9$ cm. This radius is slightly larger than that obtained in the results in [2], where $R_2$ ~ $0.047 \pm 0.001\ R_\odot$

To obtain a feedback from these results, we check the value of the donor star $M_2$ and the calculated value of $R_2$ by applying them in the formulae of [10,29,30]:

$$P_{orb}(h) = 8.75\left(M_2 / R_2^3\right)^{-1/2} \tag{4}$$

We see that the checked orbital period $P_{orb}(h)\ [check] = 0.39h$ is very close to the accepted value of the orbital period of 24.5 min (~0.40 h) [5,6]. So, the values of $M_2$ and $R_2$ fit Equation (4) well.

The set of parameters, including also the orbital period P and the super-cycle period $\tau$ [5,12,31,32], are listed in Table 1.

**Table 1.** The CR Boo system parameters, as follows: $M_1$—mass of the primary star; $M_2$—mass of the secondary star; $q$—mass ratio; $M$—the total mass; $P$—orbital period; $\tau$ (days)—super-cycle period; $R_1$—radius of the primary star; $R_2$—radius of the secondary star; $a$—the orbital separation between the components.

| $M_1$ ($M_\odot$) | $M_2$ ($M_\odot$) | q | M ($M_1 + M_2$) | P (min) | $\tau$ (d) | $R_1$ ($R_\odot$) | $R_2$ ($R_\odot$) | $a$ ($R_\odot$) |
|---|---|---|---|---|---|---|---|---|
| 0.80 | 0.07 | 0.087 | 0.87 | 24.5 | ~46 | 0.012 | 0.0526 | 0.266 |

Following the observational data in Section 2, the next step is to track the luminosity changes during our observations. Now, as we have data of apparent magnitudes for the UBV bands, for each point of time, it is easy to determine the absolute magnitude, taking into account the distance to the object (distance to CR Boo, d = 337 pc [33]) and the bolometric correction (BCv ~ −2.5 [7]). Next, we calculated a single luminosity value for each observational point, in detail, for the two nights and separately for UBV bands, by the relation between the absolute magnitude $M_{UBV}$ and luminosity $L_{UBV}$:

$$M_{UBV} = -2.5 \log\left(\frac{L_{UBV}}{L_\odot}\right) \tag{5}$$

We employed suitable mathematical software tools that consist of inbuilt methods of calculations. The results of the calculations show that the variations in luminosity values are in a range ≈ 0.004–0.008, on average, for all bands. The estimated errors are ≈ 0.0001–0.0002 for each band and the standard deviation of the values ≈ 0.0012–0.0027.

Here in Table 2, we give the minimum, average and maximum values of the luminosity.

**Table 2.** Mean values of luminosity $L$ ($L_\odot$), calculated for two days of observations for each band separately. The delimitation of V1 and V2 is only by the different dates. The order of luminosity in cells: min/mean/max values.

| Date\Band | V1 | V2 | U | B_sh | B_vid |
|-----------|--------|--------|--------|--------|--------|
| 1 July 2019 | | 0.0279 | | | |
| | | 0.0318 | | | |
| | | 0.0352 | | | |
| 5 July 2019 | 0.0154 | | 0.0372 | 0.0143 | 0.0138 |
| | 0.0171 | | 0.0406 | 0.0161 | 0.0166 |
| | 0.0194 | | 0.0461 | 0.0184 | 0.0196 |

During our observations, the object's magnitude varied between 16.1 and 17 in UBV, which is in the lower observable range, defined by the results of [4,14,15] (see also the Introduction). Then, CR Boo was in its quiescent state period—1 and 5 July 2019. From the observational journal, we first analyzed in detail each observational data point and related them to the corresponding magnitude. Then, applying a simple mathematical distribution, we defined the step by which the values change, by extracting their maximum and minimum. Thus, according to the points of time in the data, we estimate that the amplitude of our brightness variations reaches an average value of 0.2 mag.

We suggest the observed small-scale or small-amplitude variability in the magnitude, as in the luminosity (Table 2), which appeared in the V and UBV bands, are most likely a manifestation of "humps", according to the effect's definition in [6,11,18,19] (see Section 1). The humps usually appear in low states and their average amplitude is in the interval of 0.2–0.3 mag, which is close to our estimation.

Although we could now relate both effects, we calculated the hump period Ph by again applying the PDM method. The average value in the 1st observational night is a Ph of ~23.2 min in V, and a Ph of ~24 min in the 2nd night, in UBV. Thus, the coincidence of the periodicity of the humps with those calculated in Section 2 is seen as well. The period of these amplitude variations is very close to the orbital period of 24.5 min. For this reason, their behavior is also recognized as orbital humps.

In the quiescent states, the appearing of humps is usually caused by the periodical visibility of the hot spot [19,34], which is placed at the outer part of the accretion disc, where the inflow from the secondary star contacts the edge of the disc around the primary star.

## 4. Conclusions

We presented the observations of the AM CVn star CR Boo in the UBV bands, made on two nights in July 2019: 5 July 2019 and 1 July 2019. The observed magnitude was in the range of 16–17 mag in the V band, which corresponds to the object's low state. We detected small-scale variations, seen in the light curves in all bands, with an amplitude of ~0.1–0.2. The radii of the two components of CR Boo and the orbital separation $a$ were estimated by applying the accepted values of their masses. The calculated radii belong to a range of values, which correspond to the estimated radii of many white dwarf stars, usually in the order of $10^8$–$10^9$ cm. [35–38]. In our opinion, it could be a useful feature in support of CR Boo's components description.

With all this, according to the obtained values of the orbital separation and the radii of the two components, we suggest that the current state of CR Boo is following the case, when the mass transfer is in progress, with $a$ larger than $R_1$ and $R_2$, by around an order. The estimation of the mass-transfer rate is a subject of our next paper.

On the base of these observations, we calculated the luminosity values for each band. The obtained values of luminosity are consistent with the range of the white dwarf's luminosity, $L \approx 10^{-3}$–1 $L_\odot$. The observed small-amplitude brightness variations on those two nights are probably due to hump production. Since these humps are appearing during the quiescent state, we suggest that the most possible origin is the existence of a hot spot and its periodical visibility.

Most of the other authors usually pay attention to the production of superhumps during the outburst periods of the AM CVn objects. In this paper, we discuss that the humps manifested during the quiescent state of CR Boo could also be a subject of research. We show that the hump production in CR Boo could be recognized even on two nights of observations.

**Author Contributions:** Conceptualization, D.B., Z.C. and S.B.; data reduction R.Z.; observations, S.B., Y.N. and Z.C.; analysis, D.B.; writing—original draft, D.B. All authors have read and agreed to the published version of the manuscript.

**Funding:** This work is supported by the grant: Binary stars with compact objects, КП-06-Н28/2 08.12.2018 (Bulgarian National Science Fund).

**Acknowledgments:** The authors thank the referees for the useful comments. We thank to the SBACXII organizing committee for the possibility to present and publish this work.

**Conflicts of Interest:** The authors declare no conflict of interest.

**Publisher's Note:** MDPI stays neutral with regard to jurisdictional claims in published maps and institutional affiliations.